\documentclass[aps,prl,twocolumn,groupedaddress]{revtex4}

\usepackage{graphicx}
\usepackage{xspace}
\usepackage{booktabs}
\usepackage{psfig}
\usepackage{amsmath}
\usepackage{fancyhdr}

\begin{document}

\title{Enhanced surface plasmon resonance absorption \\ in metal-dielectric-metal layered microspheres}

\author{K. Hasegawa}
\author{C. A. Rohde}
\author{M. Deutsch}

\affiliation{Oregon Center for Optics and Department of Physics,
University of Oregon, Eugene, OR 97403}

\date{\today}

\begin{abstract}We present a theoretical study of the dispersion relation of
surface plasmon resonances of mesoscopic metal-dielectric-metal microspheres.
By analyzing the solutions to Maxwell's equations, we
obtain a simple geometric condition for which the system exhibits a
band of surface plasmon modes whose resonant frequencies are weakly
dependent on the multipole number. Using a modified Mie calculation,
we find that a large number of modes belonging to this
flat-dispersion band can be excited simultaneously by a plane wave,
thus enhancing the absorption cross-section. We demonstrate that the
enhanced absorption peak of the sphere is geometrically tunable
over the entire visible range.\end{abstract}

\maketitle

\thispagestyle{fancy} \fancyhead[C]{\sf Accepted for publication in
{\sl Optics Letters}}

\noindent In recent years there has been a growing interest in
surface plasmon resonances (SPRs) of metal-dielectric
structures as means to concentrate electromagnetic (EM) field in
subwavelength volumes. The geometric tunability and the enhanced
field intensity at the metal-dielectric interface have led to a
number of theoretical and experimental demonstrations of
SPR-assisted EM energy focusing\cite{Li03,Schuck05,Muhlschlegel05}.
In particular, extensive studies have been conducted on SPRs of
nanoshells---thin metal shells surrounding sub-micron
dielectric cores. Both single\cite{Jackson03} and concentric\cite{Prodan03}
nanoshell systems have been addressed.
It was shown that the resonant frequencies, where incident EM radiation is efficiently
focused to the nanoshells, are tunable from the visible to
near-infrared. Moreover, the large EM field
concentration at the metallic shell produces a giant surface
enhanced Raman scattering\cite{Jackson03}.

The size scale of a typical nanoshell is of the order of
$100\hbox{nm}$ or less, a small fraction of the excitation wavelength. As a
result, only a single, low order multipole resonance plays a vital
role in channelling the incident EM energy towards the metallic shell.
On the other hand, mesoscopic layered particles open up the
possibility for simultaneous excitation of a large number of SPRs,
further increasing the efficiency of the EM field focusing. This, in
turn, enhances optical phenomena such as absorption and nonlinear
response.

In this Letter, we show how to utilize dispersion engineering to
enhance the EM field focusing and the absorption cross-section of
mesoscopic multilayered spheres. The system consists of a spherical
resonator comprised of thin, alternating layers of dielectric and
metal shells around a concentric metal core. The composite
particle is embedded in a homogeneous, isotropic dielectric host
with permittivity $\epsilon_0$.
More specifically, we address metal-dielectric-metal (MDM)
microspheres: metallic cores comparable in size to optical
wavelengths, surrounded by one sequence of lossless dielectric shell
of thickness $L$ followed by a metal shell of thickness $T$. We show
that it is possible to obtain a band of SPRs with nearly identical resonant
frequencies (flat-dispersion band) by adjusting the above geometric
parameters. By solving Maxwell's equations using a spherical
multipole expansion, we calculate the dispersion relations
for the flat band and the resultant enhanced absorption
cross-section.

To analyze the SPRs excited in the system, we initially
address the dispersion relations of a simplified geometry in which the
outer metal shell is infinitely thick. This model is complementary
to a metal nanoshell on a dielectric core\cite{Jackson03}, the DMD. Unlike the DMD,
the presence of a properly designed \emph{dielectric}
shell in the MDM is necessary and sufficient for achieving a flat-dispersion band.

We first consider a metal characterized by a lossless Drude model,
$\epsilon_m(\omega)=\epsilon_b-\omega_p^2/\omega^2$, where $\epsilon_b$
is the contribution of inter-band transitions and $\omega_p$
is the plasma frequency. The dielectric shell has a real permittivity $\epsilon_d$.
The eigenfrequencies of the simplified MDM
system are determined by the eigenmode equation\cite{Ruppin82}
\begin{align} {0}
&=\biggl\{\eta\frac{h_l(k_dS)}{h_l(k_mS)}
-\frac{[k_dS\ h_l(k_dS)]'}{[
k_mS\ h_l(k_mS)]'}\biggr\} \nonumber\\
&\quad\quad\times\biggl\{\eta\frac{j_l(k_dR)}{j_l(k_mR)}-\frac{[k_dR\
j_l(k_dR)]'}{[
k_mR\ j_l(k_mR)]'}\biggr\}\nonumber\\
&\quad -\biggl\{\eta\frac{j_l(k_dS)}{h_l(k_mS)}
-\frac{[k_dS\ j_l(k_dS)]'}{[
k_mS\ h_l(k_mS)]'}\biggr\} \nonumber\\
&\quad\quad\times\biggl\{\eta\frac{h_l(k_dR)}{j_l(k_mR)}-\frac{[k_dR\
h_l(k_dR)]'}{[ k_mR\ j_l(k_mR)]'}\biggr\}\label{Eigen}
\end{align} where $\eta=1$ for transverse electric (TE) modes and $\eta=\epsilon_d/\epsilon_m$ for transverse magnetic (TM) modes, and $k_{m,d}=\sqrt{\epsilon_{m,d}}\ \omega/c$.
The radius of the inner metal core is $R$ and $S\equiv R+L$.
The spherical Bessel and Hankel functions of the first kind of integer order $l$
are denoted respectively by $j_l$ and $h_l$, and the prime denotes
differentiation with respect to the argument. We limit our
discussion to large enough particles, satisfying $1\ll k_dR$ and
$1\ll |k_m|R$. Using a real $\epsilon_m(\omega)$ leads to straightforward analytic
solutions of Eq. (\ref{Eigen}) with real $\omega$. We show later that using a realistic
(lossy) metal only modifies quantitative aspects of the flat band, while
its fundamental physical origins remain unaltered.

For each polarization there is an infinite number of solutions to Eq. (\ref{Eigen}), each characterized by a multipole number, $l$ and $n\geq
0$ roots, the latter yielding a radial excitation number (i.e.
\emph{band index}). We first examine the resonant modes for
asymptotic limits of  Eq. (\ref{Eigen}). For high-order multipoles
satisfying $l\gg |k_m|S$ and $l\gg k_dS$ we expand $j_l$ and $h_l$
to obtain the resonance condition
\begin{equation}
\epsilon_m(\omega_l)+\epsilon_d=0\label{lggkR}.
\end{equation} for TM polarization. No similar resonance condition exists for TE polarization.
We next look for an expression for $L$ such that Eq.
(\ref{lggkR}) is also satisfied for the TM mode with the
smallest multipole number. For $l=1$ where expansions of $j_l$ and $h_l$
may be applied, Eq. (\ref{Eigen}) reduces to
\begin{equation}
(2n-1)\pi=2k_d L,\label{lllkR0}
\end{equation} with $n\geq 1$.

The high order multipole modes of Eq. (\ref{lggkR}) belong to the
$n=1$ radial excitation band, hence by setting $n=1$ in Eq.
(\ref{lllkR0}), we obtain the geometric condition
\begin{equation}L=L^{\star}\equiv\frac{\lambda_{sp}}{4\sqrt{\epsilon_d}}\label{lllkR}\end{equation}
where $\lambda_{sp}=2\pi c/\omega_{sp}$ and $\omega_{sp}$ is given
by $\epsilon_m(\omega_{sp})+\epsilon_d=0$. The position of the flat
band given by Eq. (\ref{lggkR}) and the geometric condition, Eq.
(\ref{lllkR}), do not depend on $R$ and are identical to those for
the planar MDM geometry\cite{Shin04}.

\begin{figure}[t]
\centerline{\includegraphics[width=8.3cm]{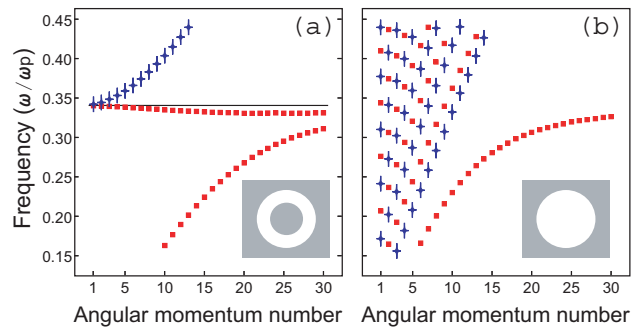}} \caption{
\label{fig1} (a) Dispersion relations of TM (squares) and TE
(crosses) modes of the MDM sphere. Near the plasmon frequency
$\omega_{sp}$ (solid line), there is a flat band of TM modes weakly
dependent on multipole number. (b) Dispersion relation of same
structure without metal core. The insets show schematic cross-sections of
the systems.}
\end{figure}

Figure \ref{fig1}(a) shows the resonant frequencies of the system plotted as function of
the multipole number $l$. To approximate the optical response of
silver, we have chosen $\epsilon_b=5.1$ and
$\hbar\omega_p=9.1\hbox{eV}$\cite{Gadenne98}.  The dielectric shell
has $\epsilon_d=3.53$, corresponding to the measured value of
amorphous titania. For clarity, we have chosen a high dielectric
constant to place $\omega_{sp}$ well away from the bulk plasma
resonance at $\omega_p$. Using Eq. (\ref{lllkR})
$L^{\star}=53.4\hbox{nm}$, and we set $R=500\hbox{nm}$. From the
figure, it is clear that there is a band of TM modes whose
frequencies are weakly dependent on the multipole number. This
flat-dispersion band is near
$\omega_{sp}/\omega_p=(\epsilon_b+\epsilon_d)^{-1/2}\approx 0.34$ as
expected, and the width of the band is given by
$\delta\omega/\omega_{sp}=0.014$ where $\delta\omega$ is the
difference of the largest and the smallest frequencies in the band.
For comparison, Fig. \ref{fig1}(b) shows the dispersion relation of
a dielectric sphere of radius $R+L$ embedded in a metallic host.

By analyzing the dispersion relation of MDM spheres with various
core radii, we have found that when $R\geq 100\hbox{nm}$,
$\delta\omega$ depends weakly on the core radius $R$. For
example, $\delta\omega/\omega_{sp}=0.018$ for $R=100\hbox{nm}$ and
$\delta\omega/\omega_{sp}=0.014$ for $R=1000\hbox{nm}$. We have
observed that the material dispersion of the metallic medium affects
the flatness $\delta\omega/\omega_{sp}$ more significantly; a larger
and positive $d\epsilon_m/d \omega|_{\omega_{sp}}$ leads to a
flatter dispersion relation.

Next, we analyze the absorption cross-section of the MDM sphere with
a metal shell of finite thickness, and study the coupling between the incident
plane wave and the flat-dispersion modes. Such spheres may be realized by chemical synthesis of layered metallodielectric particles\cite{Prodan03,Velikov03}. We have developed an exact and numerically stable algorithm for calculating the absorption cross-section of multilayered spheres based on previous publications \cite{Toon81,Kaiser93,Yang03,Du04,Cachorro91}.
We also account for absorption losses in the metal by modifying the
Drude model to
$\epsilon_m(\omega)=\epsilon_b-\omega_p^2(\omega^2+i\Gamma\omega)^{-1}$,
with $\Gamma$ describing the electron relaxation rate.
We set $\hbar\Gamma=0.021\hbox{eV}$ \cite{Gadenne98} and $\epsilon_0=1$.

\begin{figure}[t]
\centerline{\includegraphics[width=8.3cm]{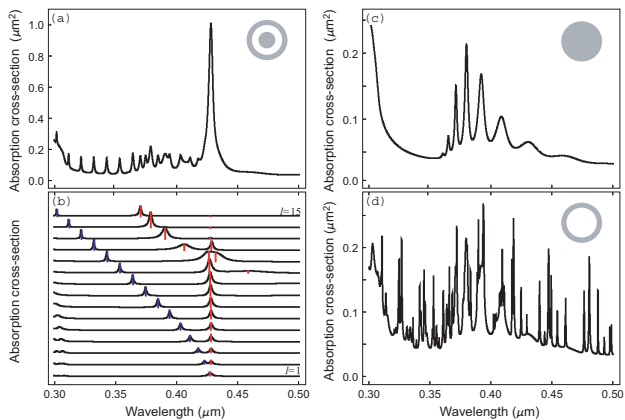}} \caption{
\label{fig2} (a) Optimized absorption cross-section spectrum of MDM
sphere with $R=500\hbox{nm}$. (b) Mode decomposition of the spectrum
from $l=1$ (bottom) to $l=15$ (top), showing contributions from TE
modes (blue hatches), and TM modes (red hatches). (c) Absorption
spectrum of a metal sphere of radius $R+L+T$. (d) Absorption
spectrum of a sphere with dielectric core of radius $R+L$ and metal
shell of thickness $T$.}
\end{figure}

A sharp absorption peak near $\lambda_{sp}$ is seen in
Fig. \ref{fig2}(a). For a core of $R=500\hbox{nm}$
this is maximized when $L=62\hbox{nm}$ and
$T=77\hbox{nm}$, and it is positioned at $\lambda=428\hbox{nm}$. The
deviations of the peak from $\lambda_{sp}=401\hbox{nm}$ and
of the optimal value of $L$ from $L^{\star}=53.4$ are
explained as follows: In addition to SPRs of the flat-dispersion
band, SPRs of the outermost metal-host interface contribute to
this peak. In fact, the two SPR branches are coupled through the
finite metal shell. These plasmon hybridizations lead to slight
modifications in peak position as well as in the optimal dielectric
shell thickness.

The degree of coupling between the incident plane wave and the flat
dispersion band is tuned by adjusting $T$, the thickness of the
outer metal shell. As we show later, the value of $L$ may be varied
in the vicinity of $L^{\star}$ to spectrally tune the absorption maximum
while maintaining a nearly flat dispersion band. For each such value of $L$
there exists a $T$ which optimizes the coupling. Nevertheless, if the
metal shell is made too thin, strong plasmon hybridization will
eventually distort the flat band.

\begin{figure}[t]
\centerline{\includegraphics[width=8.3cm]{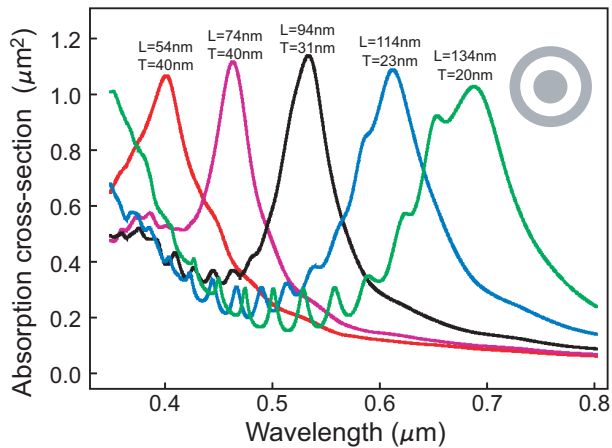}} \caption{
\label{fig3} Absorption spectra of silver-titania MDM sphere with
$R=500\hbox{nm}$ and various dielectric shell thickness $L$. The
value of $T$ is chosen to maximize the peak height. }
\end{figure}

The mode decomposition of the spectrum, shown in Fig. \ref{fig2}(b),
confirms that a large number of multipoles share nearly identical
resonant frequencies and shows that the first 13 multipoles are
excited concurrently by the incident plane wave. Higher order
multipoles have negligible contribution to the absorption peak.
Compared to the absorption spectrum of a metal sphere and of a
core-shell of the same size, plotted in Fig. \ref{fig2}(c) and (d),
the absorption cross-section of the MDM near $\lambda=428\hbox{nm}$
is enhanced by a factor of 10 and 4, respectively.
The enhancement factor relative to the maximal absorption is seen to be $\sim4$ for both cases.
The enhanced absorption exists for all
core radii $R\gtrsim 100\hbox{nm}$. The value of $R$ affects the
peak height and the number of simultaneously excited SPRs. Other
spectral characteristics, such as the peak position, its FWHM, and
the optimal values of $L$ and $T$, are insensitive to the core size.
For example, a change of a few hundred nanometers in the core radius
alters the latter by less than several nanometers. For radii smaller
than $100\hbox{nm}$, the absorption broadens, splitting into
multiple peaks. Below $R=50\hbox{nm}$, the effect of the metal core
on the absorption diminishes, the spectrum rapidly approaching that
of a DMD nanoshell.

The enhanced absorption is also demonstrated numerically using the
experimentally obtained dielectric function for
silver\cite{Palik85}. For $R=500\hbox{nm}$, the absorption
cross-section peak is maximized for $L=94\hbox{nm}$ and
$T=31\hbox{nm}$, and it is at $\lambda=533\hbox{nm}$ with
$64\hbox{nm}$ FWHM as shown in Fig. \ref{fig3}. The absorption
enhancement factors relative to a solid metallic sphere and to a
core-shell sphere of a comparable size are 7 and 3, respectively.
While the spectral position of the flat dispersion (i.e.
$\omega_{sp}$) depends only on the optical properties of the
constituent media, the absorption peak position can be tuned by
adjusting $L$. As discussed above, an arbitrary $L$ does not produce
a flat dispersion. However, the slope of the band,
$\omega_l-\omega_{l-1}$, is generally small as long as $L\sim
L^{\star}$. Thus, it is possible to vary the value of $L$ by tens of
nanometers and shift the resonant frequencies, and still retain a
significant overlap of the absorption cross-section peaks of various
multipoles. Figure 3 shows the geometric tunability of the enhanced
absorption over the visible range.

In summary, we have analyzed the dispersion relation of mesoscale
MDM spheres, and derived the condition for which the system exhibits
a band of TM modes whose eigenfrequencies are weakly dependent on
the multipole number. The numerically obtained absorption spectra
exhibit enhanced absorption as a consequence of the simultaneous
excitation of a large number of SPRs. Since the enhancement implies
a large EM field concentration, these results suggest that the
flat-dispersion MDM spheres may be used to enhance nonlinear optical
phenomena, and more generally, they may lead to a new generation of
mesoscale plasmonic systems in which numerous modes are excited to
manipulate SPR-assisted light-matter interactions.

This work was supported by NSF Grant No. DMR-02-39273 and ARO Grant
No. DAAD19-02-1-0286.

\end{document}